\documentclass[conference,compsoc]{IEEEtran}

\ifCLASSOPTIONcompsoc
  \usepackage[nocompress]{cite}
\else
  \usepackage{cite}
\fi

\ifCLASSINFOpdf
\usepackage{xcolor}
\usepackage{graphicx} 
\usepackage{listings} 
\lstdefinelanguage{JavaScript}{
  morekeywords={await,async,yield,import,from,export,default,let,const,new,return},
  sensitive=true,
  morecomment=[l]{//},
  morecomment=[s]{/*}{*/},
  morestring=[b]',
  morestring=[b]"
}

\else
\fi

\usepackage{url}

\usepackage{xcolor}
\usepackage{tcolorbox}
\usepackage{tabularx}
\usepackage{multirow}
\usepackage{booktabs}
\newtcolorbox{findingbox}{
  colback=gray!5,
  colframe=black!60,
  boxrule=0.8pt,
  arc=4pt,
  left=8pt,
  right=8pt,
  top=6pt,
  bottom=6pt
}
\newcommand{\circlednum}[1]{\tikz[baseline=(char.base)]{
            \node[shape=circle,draw,inner sep=0pt, minimum size=9pt, scale=0.8] (char) {#1};}}

\usepackage{url}
\usepackage{array}
\usepackage{tabularx}
\usepackage{multirow}
\usepackage{tikz}
\usepackage{tcolorbox}
\usetikzlibrary{shapes.geometric, arrows.meta, positioning, calc, fit, backgrounds, decorations.pathreplacing}

\usepackage{fancyhdr}

\begin{document}
%
\title{A First Measurement Study on Authentication Security \\ in Real-World Remote MCP Servers}

\author{\centering
  \begin{tabular}{c c c}
    \textbf{Huijun Zhou\textsuperscript{*}} & \textbf{Xiaohan Zhang\textsuperscript{*}} & \textbf{Haozhe Zhang} \\
    Fudan University & Fudan University & Fudan University \\
    zhouhj24@m.fudan.edu.cn & xh\_zhang@fudan.edu.cn & haozhezhang25@m.fudan.edu.cn \\[1.2ex]
    \textbf{Haoyang Zhang} & \textbf{Mi Zhang} & \textbf{Min Yang} \\
    Central South University & Fudan University & Fudan University \\
    8210230811@csu.edu.cn & mi\_zhang@fudan.edu.cn & m\_yang@fudan.edu.cn \\
  \end{tabular}
}

\maketitle

\begingroup
\makeatletter
\renewcommand\@makefntext[1]{\noindent #1}
\makeatother
\footnotetext{* Both authors contributed equally to this research.}
\endgroup

\begin{abstract}
The Model Context Protocol (MCP) is emerging as a common interface connecting large language models (LLMs) with external services. 
Remote deployments are becoming increasingly important as agents connect to user-linked online services, such as social, productivity, and financial services. 
In such deployments, the authentication boundary between MCP clients and remote servers becomes security-critical, yet remains underexplored.

We present the first measurement study of authentication security in real-world remote MCP servers.
We identify 7,973 live remote MCP servers, finding that 40.55\% expose tools without authentication. 
Among authenticated servers, OAuth is the dominant authorization mechanism for reaching remote services, and OAuth deployments in the MCP ecosystem commonly exhibit three characteristics: \textit{open client environments}, \textit{dynamic client registration}, and \textit{delegated authorization}.
These characteristics distinguish MCP deployments from traditional OAuth and introduce new attack surfaces.
Guided by this observation, we derive a taxonomy of authentication flaws comprising three MCP-specific categories and conventional OAuth misconfigurations, for a total of four categories and nine concrete flaw types.
To evaluate these flaws at scale, we implement a semi-automated detection framework that combines passive traffic inspection with active dynamic probing. 
Applying it to 119 testable real-world OAuth-enabled MCP servers, we find that each server exhibits at least one flaw, with a total of 325 flaws identified, among which dynamic client registration flaws affect 96.6\% of tested servers.
Many of these flaws can lead to sensitive information leakage and account takeover. 
Through responsible disclosure, we obtained 9 CVE IDs. 
Our findings expose pervasive authentication weaknesses in the MCP ecosystem and underscore the urgent need for hardened OAuth-based remote deployments.

\end{abstract}


\IEEEpeerreviewmaketitle

\section{Introduction}
Large language models (LLMs) are increasingly used as agents that invoke tools, retrieve data, and act on external systems. This shift requires a reusable interface for agents to discover tools, exchange context, and invoke operations on behalf of users. The Model Context Protocol (MCP) has emerged as such an interface, allowing LLM clients such as Claude Desktop, Cursor, and terminal-based agent frameworks to connect to third-party servers that expose files, databases, APIs, and other capabilities.

MCP servers can be deployed locally or remotely. Local servers run on the user's device and are a natural fit for exposing device-local resources within a default-trusted environment.
Anthropic's recent deployment guidance suggests a shift in MCP's role: local capabilities can often be handled through application-native skills or CLI-based agent workflows, while MCP is positioned as the interface for agents to reach production systems and remote services, especially for Web applications, mobile apps, and cloud-hosted agents~\cite{anthropic2026production}. As a result, remote MCP servers become the service endpoints through which different MCP clients can access user-linked online accounts and trigger privileged operations, including invoking SaaS APIs, querying private data, and manipulating cloud resources~\cite{ligao2025mcp, yang2026compatibility}.

\begin{figure}[!t]
    \centering
    \includegraphics[width=\linewidth]{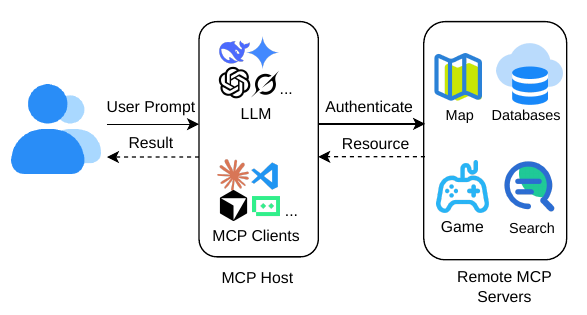}
    \caption{Demonstration of the remote MCP server authentication.}
    \label{fig:mcp-overview}
\end{figure}

Figure~\ref{fig:mcp-overview} illustrates the security setting studied in this paper. A user may ask an agent to interact with remote services through an MCP client, and the client in turn connects to remote MCP servers that expose service-backed tools. These tools may be linked to the user's social platforms, maps, entertainment applications, cloud workspaces, payment services, or other remote accounts. In this setting, authentication forms the security boundary before an MCP client can access server-side tools and the account-linked capabilities behind them. If this boundary is missing or incorrectly enforced, an attacker-controlled client may gain unauthorized access to the user-linked service accounts.

Securing this boundary is challenging because MCP brings OAuth into a loosely coupled agent ecosystem. 
MCP clients and servers, and upstream authorization servers are often developed and operated by different entities. Some deployments further introduce a cross-entity, multi-hop authorization model in which the MCP server acts as an intermediary between local clients and upstream services~\cite{obsidian2026mcp}. Although the MCP specification adopts OAuth-based authorization and outlines best practices~\cite{mcpspec2025stable}, real-world implementations may deviate from standard OAuth flows, a pattern that has repeatedly led to vulnerabilities in prior OAuth systems~\cite{philippaerts2022oauch}. Existing MCP security research has mainly studied model-layer and tool-interface threats, such as prompt injection~\cite{guo2025mcp, radosevich2025mcp} and tool poisoning~\cite{hasan2025mcp, zhao2025taxonomy}, leaving protocol-layer authentication security less understood.

To address these gaps, we conduct the first large-scale measurement of authentication security in real-world remote MCP servers. We first characterize the deployment landscape of remote MCP servers and analyze their authentication adoption and OAuth practices. We then examine how OAuth is deployed in this setting and uncover the implementation deviations.
Using mainstream cybersecurity search engines, we identify 7,973 validated live remote MCP servers. Among them, 40.55\% expose their tool interface with no authentication mechanism, while 2,428 implement OAuth-based authorization flows. 
We then analyze a fully testable subset of these OAuth deployments and distill three security-relevant characteristics: \textit{open client environments}, \textit{dynamic client registration}, and \textit{delegated authorization}.

Based on these observations, we construct an authentication flaw taxonomy for OAuth-based remote MCP, summarizing 9 security flaws across 4 categories. 
We also develop a semi-automated detection framework that reconstructs OAuth lifecycles from MCP traffic, applies passive checks to observed flows, and performs controlled active probing for flaws that require dynamic validation. 
Applying it to 119 OAuth-enabled MCP servers, we find that each tested server exhibits at least one flaw, with a total of 325 flaws identified.
Among them, dynamic client registration flaws and open client environment flaws affect 96.6\% and 85.7\% of servers respectively.
Many of these flaws can lead to sensitive information leakage and account takeover. 
We responsibly disclosed confirmed issues to affected vendors and obtained 9 CVE IDs.
The main contributions of this paper are as follows:

\begin{itemize}
\item \textbf{First Measurement of Remote MCP Authentication.}
We present the first empirical analysis of the deployment scale, authentication adoption, and OAuth authorization practices of remote MCP servers in the wild.

\item \textbf{Authentication Security Analysis.}
We perform an in-depth analysis of OAuth-based remote MCP authentication flows, distill three architectural characteristics that shape their attack surface, and propose a flaw taxonomy comprising 4 categories and 9 distinct flaws grounded in these characteristics.

\item \textbf{Real-world Evaluation.}
We develop a semi-automated security detection framework and conduct a large-scale detection study of OAuth-based remote MCP servers. 
We manually tested 119 OAuth-enabled servers and found that all exhibit at least one flaw, obtaining 9 CVE IDs through responsible disclosure. 
\end{itemize}

\section{Background}

\subsection{Model Context Protocol (MCP)}

The Model Context Protocol (MCP), introduced by Anthropic in November 2024, is an open standard for connecting LLM-based applications to external tools, data sources, and services~\cite{mcp}. Its architecture consists of three principal roles: 
(1) The \textit{MCP host} is the user-facing AI application, such as Claude Desktop, Cursor IDE, or an autonomous AI agent, that orchestrates LLM interactions, enforces access control, and manages the lifecycle of MCP client connections. 
(2) The \textit{MCP client} is an intermediary embedded within the host that manages bidirectional communication with one or more MCP servers. It initiates capability discovery requests, dispatches tool invocations, and processes server notifications and responses. 
(3) The \textit{MCP server} is an independent process that exposes three categories of capabilities to clients: \textit{tools}, \textit{resources}, and \textit{prompts}~\cite{errico2025securing, maloyan2026breaking}. 
All messages between MCP clients and servers are encoded as JSON-RPC 2.0 over UTF-8.

MCP defines two standard transport modes. In the stdio transport, the MCP host launches the MCP server as a local subprocess and exchanges JSON-RPC messages over standard input and output streams. In the HTTP/SSE transport (Streamable HTTP), the MCP server runs as a network-accessible process and communicates via HTTP POST requests and Server-Sent Events~\cite{hasan2025mcp, gaire2025sok}. 
This distinction determines the basic security boundary of the deployment. In the stdio model, the MCP server remains within the same trust boundary as the host. 
In the HTTP/SSE model, the MCP server is exposed as a network service and therefore inherits the authentication, access-control, and transport security requirements of a conventional Web application~\cite{ligao2025mcp, yang2026compatibility}.

\subsection{MCP Authentication}

\begin{figure}[htbp]
    \centering
    \includegraphics[width=\linewidth]{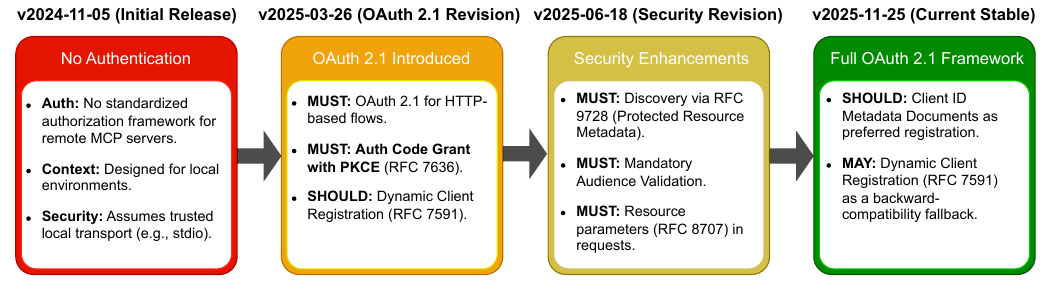}
    \caption{MCP specification authentication evolution timeline. 
    }
    \label{fig:oauth evolution}
\end{figure}

The MCP specification has evolved rapidly in its treatment of authentication for remote deployments. Figure~\ref{fig:oauth evolution} illustrates this progression.

The initial specification release on 2024-11-05 contained no mandatory authentication requirements for remote MCP servers, reflecting MCP's origins as a local integration tool~\cite{mcpspec20241105}.
The 2025-03-26 revision introduced OAuth 2.1 as the mandatory framework for HTTP-based authentication flows~\cite{mcpspec20250326}. It required authorization servers to implement OAuth 2.1 and required MCP clients to use the \textit{Authorization Code} grant with PKCE (RFC~7636)~\cite{rfc7636}.
\textit{Dynamic Client Registration} (RFC~7591)~\cite{rfc7591} was introduced as a SHOULD-level recommendation.
The 2025-06-18 revision~\cite{mcpspec20250618} further strengthened the authorization architecture by introducing \textit{Protected Resource Metadata} (RFC~9728)~\cite{rfc9728} for authorization server discovery, requiring support for OAuth 2.0 Resource Indicators (RFC~8707)~\cite{rfc8707}, and mandating audience validation.
The stable 2025-11-25~\cite{mcpspec2025stable} release further refined the client trust model by prioritizing \textit{Client ID Metadata Documents} as the preferred client registration mechanism, while retaining \textit{Dynamic Client Registration} (RFC~7591) primarily as a backward-compatibility fallback. This rapid evolution from locally trusted integrations to a more comprehensive OAuth-based security architecture may help explain the uneven levels of security compliance observed in real-world MCP deployments.

Because MCP clients are often native or locally running applications, they typically operate in open client environments and cannot reliably protect a \texttt{client\_secret}. 
In this setting, the \textit{Authorization Code} grant with \textit{PKCE} provides the main binding between the authorization request and the subsequent token exchange.
PKCE requires the client to generate a one-time \texttt{code\_verifier} and send only its derived \texttt{code\_challenge} in the authorization request. 
The original verifier is later presented at the token endpoint, so that an intercepted authorization code cannot be redeemed without possession of the corresponding verifier.

In addition, an MCP server may act as an \textit{OAuth Resource Server} with respect to the MCP client while also acting as an \textit{OAuth Client} to external services such as GitHub, Slack, or databases~\cite{obsidian2026mcp, huang2026caller}. 
This dual role creates a multi-hop authorization chain that differs from traditional two-party OAuth deployments and complicates end-to-end reasoning about security guarantees~\cite{south2025delegation, prakash2026aip}.

\subsection{Threat Model}

We follow the standard Web and OAuth security model and focus on remote attackers. The protected assets are the user's MCP session, authorization artifacts such as codes and access tokens, and the user-linked service accounts that a remote MCP server can reach. We assume that the remote MCP server, authorization server, upstream services, TLS channels, and the user's browser and device are not compromised. The security boundary of interest is therefore the authentication and OAuth flow that decides whether a particular MCP client should be allowed to access MCP tools and the account-linked capabilities behind them.

Specifically, we assume a standard attacker model. The attacker is not required to operate a malicious MCP client. Instead, they are capable of sending arbitrary HTTP requests to publicly exposed server endpoints (e.g., exploiting Dynamic Client Registration), hosting malicious web pages, and luring victims into interacting with crafted authorization URIs. The attacker can observe all network traffic directed to domains under their control (e.g., rogue redirect URIs). The attacker's goal is to bypass authentication checks, extract authorization artifacts (such as codes or tokens), bind a victim's service account to an attacker-controlled identity, or exploit the MCP server's dual role to launch confused deputy attacks against upstream resources. We do not assume that the attacker can compromise TLS cryptography, breach the victim's underlying device or browser, directly compromise the remote servers, or access non-routable enterprise-internal networks.

\section{Measurement of Remote MCP Servers}
\label{sec:measurement}

In this section, we measure how remote MCP servers are deployed in the wild. We first identify candidate servers through search-engine discovery and active probing, then summarize the resulting dataset, and finally characterize OAuth deployments in the subset that we analyze in depth.

\subsection{Identifying Remote MCP servers}

\begin{figure}[!h]
    \centering
    \includegraphics[width=\columnwidth]{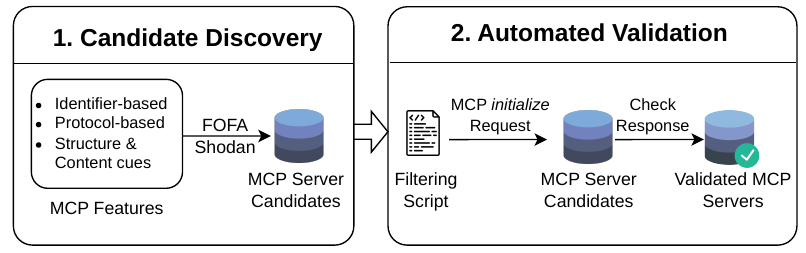}
    
    \caption{Two-step pipeline for discovering and validating remote MCP servers.}
    \label{fig:pipeline}
\end{figure}

We designed a two-step pipeline to identify remote MCP servers in the wild, illustrated in Figure~\ref{fig:pipeline}. The pipeline first collects candidates from cybersecurity search engines using MCP-specific fingerprints, and then actively probes each candidate with MCP handshake requests to confirm whether it behaves as a live remote MCP server.

\textit{(1) Candidate Discovery.}
We used two popular cybersecurity search engines, i.e., FOFA~\cite{fofa} and Shodan~\cite{shodan}, to collect candidate remote MCP servers, following prior Internet-wide measurement studies that rely on search-engine-backed asset discovery~\cite{bennett2021censysshodan}. 
Our queries combined identifier-based signals (e.g., \textit{mcp-session-id}, \textit{mcp-version}, and MCP-related hostnames), protocol-level strings (e.g., \textit{tools/call}, \textit{tools/list}, and \textit{initialize} payloads containing \textit{jsonrpc}), and lightweight structure and content cues from non-HTML endpoints such as \textit{/info}. 
We explicitly excluded conventional frontend Web features such as \textit{text/html} to reduce noise from ordinary websites.

\textit{(2) Automated Validation.}
We developed an active probing script to verify the candidate MCP servers. 
It sends an MCP \textit{initialize} handshake request and retains a node only if it returns a structurally valid JSON-RPC response with MCP protocol features. 
The script also extracts transport modes, capability configurations, authentication behavior, and basic metadata for subsequent analysis.

\subsection{Identification Results}
\label{subsec:identification_result}

Table~\ref{tab:identification-summary} summarizes the identification results. The search-engine queries produced a large initial dataset, which resulted in 28,715 unique candidate endpoints after deduplication by IP address and port. Active MCP probing further refined this set to 7,973 live remote MCP servers. To estimate false positives, two security researchers independently reviewed a random sample of 100 validated servers. We identified only 1 false positive, a non-standard JSON-RPC service that resembled an MCP handshake but did not expose a valid MCP tool interface.

\begin{table}[h]
\centering
\caption{Identification results for remote MCP servers.}
\label{tab:identification-summary}
\renewcommand{\arraystretch}{1.2}
\begin{tabular}{cc} 
\hline
\textbf{Candidate Discovery} & \textbf{Automated Validation} \\
\hline
28,715 & 7,973 \\
\hline
\end{tabular}
\end{table}

\textbf{Finding 1.1: Authentication mechanisms vary across validated remote MCP servers.}
Table~\ref{tab:auth-status} summarizes the authentication status of the validated servers. 3,233 (40.55\%) of the validated servers expose tool interfaces with no authentication at all, meaning that any client can invoke tools or trigger API requests without presenting credentials.

\begin{table}[h]
\centering
\caption{Authentication status among validated remote MCP servers.}
\label{tab:auth-status}
\small
\renewcommand{\arraystretch}{1.2}
\begin{tabular}{l l l}
\hline
\textbf{Authentication status} & \textbf{Servers} & \textbf{Proportion} \\
\hline
No authentication & 3,233 & 40.55\% \\
OAuth-based authentication & 2,428 & 30.45\% \\
Static token or API key & 2,312 & 29.00\% \\
\hline
\textbf{Total} & 7,973 & 100\% \\
\hline
\end{tabular}
\end{table}

Among the remaining authenticated servers, static tokens or API keys and OAuth-based flows are the two dominant mechanisms. Specifically, 2,312 (29\%) servers rely on static tokens or API keys, while 2,428 (30.45\%) servers implement OAuth-based authentication flows.

\textbf{Finding 1.2: Unauthenticated MCP servers can expose sensitive data.}
To understand the impact of unauthenticated deployments, we randomly selected MCP servers that exposed tool interfaces without authentication. Most of them were test or demonstration deployments and did not appear to contain sensitive data. However, we found one MCP server (**\footnote{Anonymized for ethical considerations.}) that exposed real sensitive information through unauthenticated tool access. This MCP server was primarily used for CRM (i.e., a customer relationship management system that stores detailed customer records, sales history, service requests, and communication logs). The server was intended to be an internal service, but mistakenly lacked any authentication mechanism. As a result, any user able to connect to this MCP server could query over 5,000 internal enterprise records, including customer names, email addresses, phone numbers, and physical addresses. We reported the issue to the affected party and obtained a CVE ID (CVE-2025-61510).

In addition, we will continue to validate other servers that expose tool interfaces without authentication. This will be part of our future work, and we will responsibly report any discovered issues to all relevant parties.

\textbf{Finding 1.3: OAuth is the main standardized mechanism for linking remote MCP servers to user service accounts.}
Static tokens and API keys are often used in single-user or manually provisioned deployments, but they do not provide a standard workflow for user login, consent, and account linking across different MCP clients. OAuth, by contrast, is designed for multi-user account linking and delegated access to remote services. We therefore focus the rest of the paper on OAuth-enabled MCP deployments, where authentication decisions directly govern access to user-linked remote service accounts.

\begin{findingbox}
\textbf{Finding 1.} Authentication practices in real-world remote MCP deployments remain uneven: many servers expose tools without authentication, and the authenticated deployments mainly rely on static tokens/API keys or OAuth. OAuth is especially important for remote MCP because it provides the standardized path for linking MCP clients to user service accounts.
\end{findingbox}

\subsection{Characterization of MCP OAuth Deployments}
\label{subsec:characterization}

To understand how OAuth is deployed in practice, we further characterize the OAuth-enabled remote MCP servers. 
Table~\ref{tab:oauth-subset} summarizes the subset construction. 
Since non-dynamic registration (i.e., manual client registration) cannot be automated in subsequent detection and requires per‑server manual handling with non‑scalable overhead, we focus our subsequent detection primarily on servers that support DCR (i.e., no manual registration).
We first probed OAuth metadata endpoints, such as \texttt{/.well-known/oauth-authorization-server} and OpenID Connect metadata endpoints, and treated the presence of a \texttt{registration\_endpoint} as evidence of DCR support. This process identified 1,118 DCR-enabled servers from 2,428 OAuth-enabled servers.

\begin{table}[h]
\centering
\caption{OAuth-enabled remote MCP servers used for evaluation.}
\label{tab:oauth-subset}
\small
\renewcommand{\arraystretch}{1.2}
\begin{tabular}{l c c c}
\hline
\textbf{Subset} & \textbf{OAuth-enabled} & \textbf{DCR-enabled} & \textbf{Testable} \\
\hline
\textbf{Servers} & 2,428 & 1,118 & 119 \\
\hline
\end{tabular}
\end{table}

To ensure valid and safe end-to-end testing, we manually filtered the initial deployments. We excluded 387 redundant nodes (e.g., domain/IP overlaps, multi-instance deployments) and 32 invalid cases that required no authentication. Furthermore, 573 servers were deemed untestable: 50 exposed did not support DCR (returning HTTP 404), 207 suffered connection or execution failures, and 316 were restricted by strict enterprise access controls. We then eliminated 7 anomalous nodes that immediately triggered a callback response upon receiving an authorization request without any user interaction, thus failing to constitute a complete OAuth semantic flow. 
These untestable factors are objective constraints (e.g., corporate network policies, server flaws, or lack of user interaction) that cannot be overcome by any automated or manual means.
Finally, we obtained a core testable subset of 119 servers. All subsequent statistical percentages are strictly scoped to this 119-server evaluation dataset.

\textbf{Finding 2.1: Dynamic client registration is common among OAuth-enabled remote MCP servers.}
Among the 2,428 OAuth-enabled servers, 1,118 advertise a \texttt{registration\_endpoint}, indicating DCR support. This accounts for 46.0\% of OAuth-enabled servers. DCR appears frequently because remote MCP servers need to support heterogeneous MCP clients, including desktop applications, IDEs, CLI tools, and cloud-hosted agents, for which manual pre-registration of every client instance is difficult to scale. DCR therefore becomes a practical registration mechanism for remote MCP OAuth deployments.

\begin{table}[h]
\centering
\caption{Prevalence of three common characteristics across 119 OAuth-enabled remote MCP servers.}
\label{tab:char-prevalence}
\small
\renewcommand{\arraystretch}{1.2}
\begin{tabular}{l c c}
\hline
\textbf{Characteristic} & \textbf{Servers} & \textbf{Proportion} \\
\hline
Open client environments    & 119/119 & 100\% \\
Dynamic client registration & 119/119 & 100\% \\
Delegated authorization     & 81/119 & 68.07\% \\
\hline
\end{tabular}
\end{table}

\textbf{Finding 2.2: All of the testable OAuth deployments run in open client environments.}
All 119 testable OAuth deployments interact with MCP clients that run in end-user or otherwise externally controlled environments, such as desktop applications, IDEs, CLI tools, browser-integrated clients, or cloud-hosted agent frontends. In these open client environments, clients cannot reliably protect a \texttt{client\_secret}. 
As a result, the security of the OAuth flow depends heavily on runtime protections such as PKCE, redirect URI binding, short-lived authorization artifacts, and correct callback handling.

\textbf{Finding 2.3: Delegated authorization is prevalent in testable OAuth deployments.}
Among the 119 testable OAuth deployments, 81 (68.07\%) integrate with external platforms or other remote services through delegated authorization. In such deployments, the MCP server acts as an OAuth resource server with respect to the MCP client, while also acting as an OAuth client to the upstream service. This creates a multi-hop authorization chain across independently operated entities and introduces additional state that must remain bound across layers.

\begin{findingbox}
\textbf{Finding 2.} 
OAuth-enabled remote MCP deployments commonly combine open client environments, dynamic client registration, and delegated authorization. 
These characteristics distinguish MCP OAuth from conventional Web OAuth deployments and shape the attack surface.
\end{findingbox}

\section{Security Analysis of MCP OAuth}
\label{sec:analysis}

In this section, we first abstract the OAuth workflow used by remote MCP deployments and identify the security checks that should be enforced at each phase. 
We then use this workflow to organize the implementation flaws observed in practice into a taxonomy that guides our later detection study.

\subsection{MCP OAuth Workflow}
\label{subsec:workflow}

\begin{figure*}[t]
    \centering
    \includegraphics[width=0.95\linewidth]{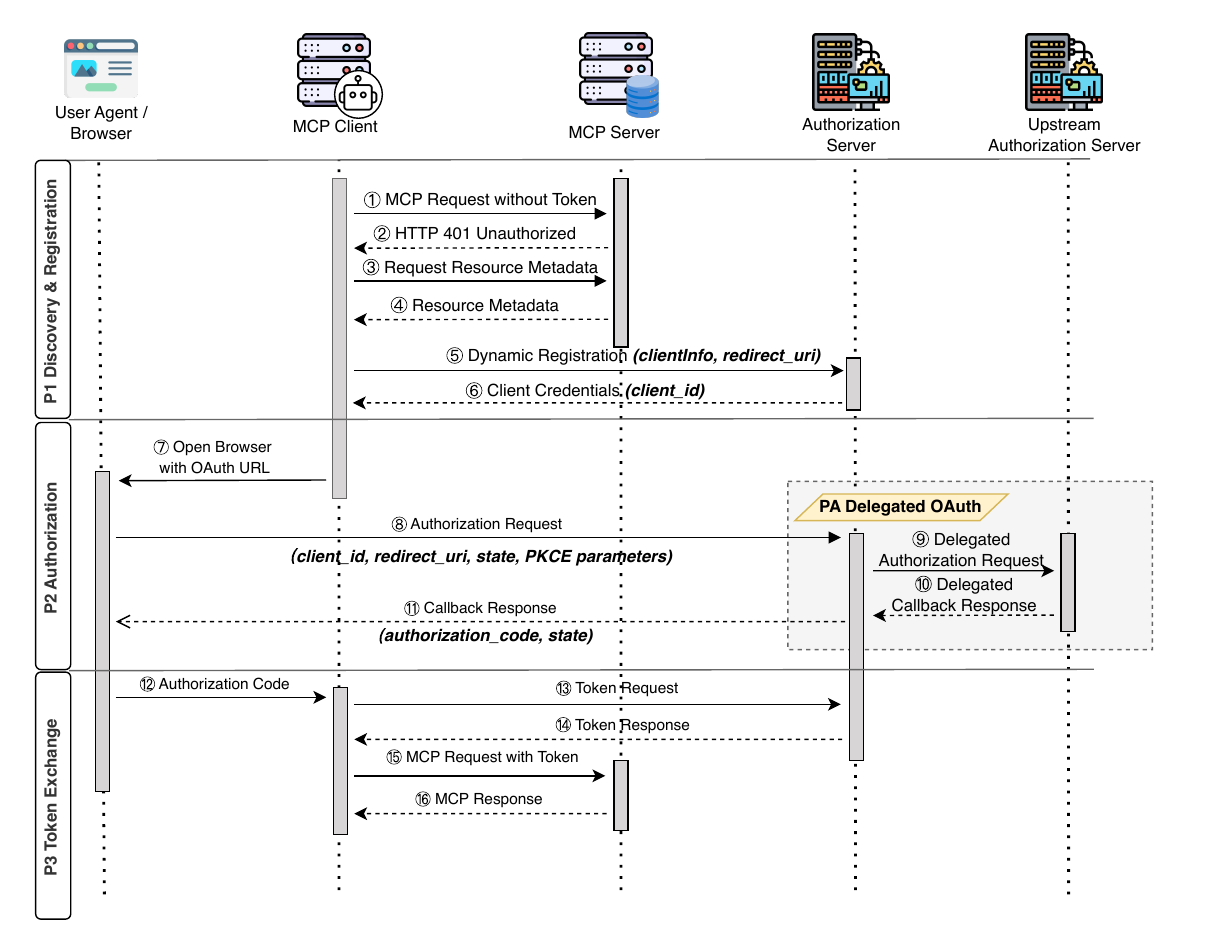}
    \caption{Workflow of OAuth-based authentication in remote MCP deployments. P1 - P3 capture the MCP client-to-server flow, while PA captures delegated authorization to upstream authorization services.}
    \label{fig:oauth-workflow}
\end{figure*}

We begin by abstracting the common OAuth workflow used by remote MCP deployments, which consists of three core phases: 
\textit{P1 Discovery \& Registration}, \textit{P2 Authorization}, and \textit{P3 Token Exchange}, plus an optional delegated phase (PA) that appears in 68.07\% of our 119-server characterization subset.
This abstraction is useful because most flaws we study correspond to a broken binding in one of these phases: between a client identity and its callback endpoint, between an authorization request and a browser callback, between an authorization code and a token exchange, or between the MCP-layer context and an upstream OAuth flow. 
Figure~\ref{fig:oauth-workflow} summarizes the abstracted workflow.

\textbf{P1: Discovery \& Registration.}
When an MCP client first accesses a protected resource on an MCP server without a token, the server returns an HTTP 401 \textit{Unauthorized} response together with metadata that points to the corresponding authorization server (\circlednum{1} - \circlednum{4}). The client then establishes a usable identity through one of the mechanisms allowed by the specification, such as \textit{Client ID Metadata Documents}, \textit{Dynamic Client Registration}, or \textit{pre-registration} (\circlednum{5} -\circlednum{6}). 
The security-critical state established in this phase is the association between a \texttt{client\_id}, its allowed \texttt{redirect\_uri} values, and the authorization server that will later issue codes. 
In remote MCP, this association is often created dynamically, so registration endpoints become part of the attack surface rather than a purely administrative interface.

\textbf{P2: Authorization.}
Once the client has the authorization endpoint and a usable identity, it constructs an authorization URL and launches the user's browser (\circlednum{7}). 
This request carries the parameters that bind the user-facing authorization decision to the client and callback context, including \texttt{client\_id}, \texttt{redirect\_uri}, \texttt{state}, and PKCE parameters. 
The authorization server authenticates the user and presents the consent interface (\circlednum{8}). 
At this point, the server should verify the client identity, enforce the registered redirect URI, preserve CSRF protection through \texttt{state}, and display enough information for the user to understand where the authorization result will be delivered.

\textbf{P3: Token Exchange.}
After successful authorization, the authorization server returns a callback carrying an authorization code and the original \texttt{state} value (\circlednum{11}). 
The browser delivers this code back to the local MCP client, which exchanges it for an access token and then includes that token in subsequent JSON-RPC requests to the MCP server (\circlednum{12} -\circlednum{14}). 
This phase should bind the authorization code to the same client and PKCE verifier used in P2, and the code should become invalid immediately after redemption. Otherwise, an attacker who obtains a code through an earlier redirect or state-handling flaw may still be able to redeem or replay it.

While P1 - P3 form a complete OAuth flow for the MCP session, many deployments also integrate third-party services that expose their own OAuth-protected APIs. 
For example, a Notion MCP server not only authenticates the MCP client through P1 - P3, but also obtains a Notion token to call the Notion API on the user's behalf. This second authorization loop is common, as 68.1\% of the OAuth-enabled servers in our 119-server subset implement delegated authorization. 
We represent this additional step as \textit{PA}.

\textbf{PA: Delegated Authorization.}
In this second-hop flow, the MCP server acts as an OAuth client to an upstream service, typically with a pre-registered \texttt{client\_id}, and manages delegated authorization on the user's behalf (\circlednum{9} - \circlednum{10}). 
This phase introduces a second authorization context that must remain consistent with the first. 
The MCP server may need to carry routing \textit{state} between the local MCP flow and the upstream OAuth flow, but that \textit{state} must be integrity-protected and bound to the correct user session. Otherwise, a flaw in the upstream context can propagate back into the MCP session.

Taken together, the four phases capture the OAuth lifecycle most commonly seen in remote MCP deployments. 
The three core phases follow the familiar OAuth 2.1 pattern, but the multi-party nature of MCP adds coordination requirements between independently operated clients, MCP servers, authorization servers, and upstream services. 
We use this workflow as the basis for the taxonomy below: each flaw corresponds to a missing or weakened check at a specific phase, and the later detection framework mirrors the same lifecycle by reconstructing these phases before applying flaw-specific tests.

\subsection{Taxonomy of Implementation Flaws}
\label{sec:taxonomy}

Guided by the abstracted workflow and the MCP and OAuth specifications, we derive a taxonomy of implementation flaws in OAuth-based remote MCP servers. 
The taxonomy asks which security checks should hold in each phase, then groups the ways these checks fail in practice. 
It contains nine flaw types in four categories: \textit{dynamic client registration flaws}, \textit{delegated authorization flaws}, \textit{open client environment flaws}, and \textit{common OAuth misconfigurations}. 

The first three categories arise from the deployment characteristics, while the last captures conventional OAuth mistakes that remain prevalent in MCP deployments. Table~\ref{tab:taxonomy} summarizes the categories and the phases where they appear.

\begin{table*}
\centering
\caption{Taxonomy of implementation flaws in OAuth-based remote MCP servers.}
\label{tab:taxonomy}
\small
\begin{tabularx}{\textwidth}{p{2.6cm} l X l}
\hline
\textbf{Category} & \textbf{Flaw} & \textbf{Description} & \textbf{Phase} \\
\hline
\multirow{2}{2.6cm}{C1: Dynamic Client Registration Flaws}
  & F1: Malicious DCR Binding & Malicious \texttt{redirect\_uri} registration via open endpoints. & P1--P2 \\
  & F2: Blind Client Trust & \texttt{client\_id} spoofing due to inadequate verification. & P2 \\
\hline
\multirow{2}{2.6cm}{C2: Delegated Authorization Flaws}
  & F3: Layer Inconsistency & Inconsistent PKCE enforcement across architectural layers. & PA \\
  & F4: Nested Context Pollution & Hijacking codes via nested \texttt{redirect\_uri} manipulation in \texttt{state}. & PA \\
\hline
\multirow{2}{2.6cm}{C3: Open Client Environment Flaws}
  & F5: PKCE Downgrade & Missing or weakened PKCE enforcement. & P2--P3 \\
    & F6: Consent Page Bypass & Missing consent display enforcement. & P2 \\
\hline
\multirow{3}{2.6cm}{C4: Common OAuth Misconf.}
  & F7: Open Redirect & Insufficient \texttt{redirect\_uri} validation. & P2 \\
  & F8: Weak State & Missing or predictable \texttt{state} enables CSRF. & P2 \\
  & F9: Code Replay & Reusable authorization code after login. & P3 \\
\hline
\end{tabularx}
\end{table*}

\smallskip\textbf{C1: Dynamic client registration flaws.}
These flaws arise when dynamic client registration accepts new clients without sufficient identity checks or parameter restrictions, allowing attackers to register malicious callbacks or impersonate trusted applications.

\textit{F1: Malicious DCR Binding.} Some authorization servers expose DCR endpoints that accept arbitrary \texttt{redirect\_uri} values from anonymous requesters. An attacker can therefore register a client bound to an attacker-controlled callback, obtain a legitimate \texttt{client\_id}, and then use it in a deceptive authorization request. If a victim completes the flow, the authorization server returns the code to the attacker's registered endpoint.

\textit{F2: Client Blind Trust.} This flaw appears when the authorization server accepts a supplied \texttt{client\_id} without verifying that it has actually been registered. An attacker can then craft an authorization request that claims the identity of a familiar application, causing the consent page to display misleading client information and increasing the chance that the user authorizes a malicious program.

\smallskip\textbf{C2: Delegated authorization flaws.}
These flaws arise in the multi-hop authorization structure where an MCP server intermediates between the local client and upstream services, creating opportunities for cross-layer policy drift and context manipulation.

\textit{F3: Layer Inconsistency.} In some delegated flows, the first-hop MCP authorization requires PKCE, but the upstream request issued by the MCP server to the upstream authorization server does not. This breaks the request-to-token binding that PKCE is meant to preserve and weakens the end-to-end guarantees of the overall flow.

\textit{F4: Nested Context Pollution.} Some MCP servers encode downstream routing state, such as a \texttt{redirect\_uri}, directly inside the upstream OAuth \texttt{state} parameter. If that nested value is neither integrity-protected nor checked against an allowlist after decoding, an attacker can tamper with the routing context and cause the authorization code to be delivered to an attacker-controlled endpoint.

\smallskip\textbf{C3: Open client environment flaws.}
These flaws stem from open client environments: MCP clients run in user-controlled environments, cannot safely store long-term secrets, and therefore rely heavily on runtime protections such as PKCE and explicit user consent.

\textit{F5: PKCE Downgrade.} MCP clients cannot rely on a protected \texttt{client\_secret} in open-client environments, so PKCE becomes the primary binding between the authorization request and the token exchange~\cite{mcpspec2025stable}. 
Failures arise when an authorization server accepts authorization requests without a \texttt{code\_challenge}, or permits the insecure \texttt{plain} method. In each case, an attacker who intercepts the authorization redirect may be able to exchange the code without possessing the expected verifier. Prior work has shown that such downgrade behavior is common in the broader OAuth ecosystem~\cite{philippaerts2022oauch}.

\textit{F6: Consent Page Bypass.}
Since MCP clients typically operate in open environments and rely on localhost callbacks, they are vulnerable to localhost impersonation. To mitigate this, the latest MCP specification mandates that authorization servers ``SHOULD display additional warnings for localhost-only redirect URIs.'' 
However, we found that certain MCP servers fail to display the \texttt{redirect\_uri}. This omission allows attackers to deceive users into unknowingly approving malicious requests, leaking authorization codes to attacker-controlled ports.

\smallskip\textbf{C4: Common OAuth misconfigurations.}
These flaws are not specific to MCP, but they remain common in MCP deployments and can combine with the MCP-specific categories above to increase impact.

\textit{F7: Open Redirect.} If the authorization server does not strictly validate \texttt{redirect\_uri} against the registered value, an attacker can substitute a malicious callback and receive the authorization code directly after the victim completes the flow.

\textit{F8: Weak State.} If the \texttt{state} parameter is missing, fixed, or predictable, the client loses its CSRF protection. An attacker can then forge an authorization request and cause the client to bind the victim's session to attacker-chosen authorization state or codes.

\textit{F9: Code Replay.} Authorization codes are meant to be single-use artifacts. If a server fails to invalidate a code immediately after redemption, an attacker who obtains that code can replay it to obtain additional access tokens.

Overall, this analysis connects MCP's deployment characteristics to concrete authentication checks in the OAuth lifecycle. The resulting taxonomy covers flaws in client registration, authorization, token exchange, and delegated authorization, and maps each flaw to the phase where evidence should appear. This phase-level structure provides the basis for our security analysis.

\section{Real-world Evaluation}
\label{sec:detection}

In this section, we examine how the flaw taxonomy manifests in real-world OAuth deployments. The detector follows the abstracted workflow: it first reconstructs the relevant OAuth lifecycle, then applies flaw-specific checks at the phases where the taxonomy indicates a security property should hold. We first explain the design ideas behind the detector, then describe the detection pipeline, report results on 119 testable OAuth-enabled remote MCP servers, and present case studies showing how individual flaws can compose into end-to-end attacks.

\subsection{Design Ideas}

Detecting MCP OAuth flaws requires more than applying independent rules to individual HTTP requests. 
OAuth traffic in MCP deployments is often mixed with local callbacks, remote MCP callbacks, and upstream service authorization flows. 
Some flaws are only visible after reconstructing a full authorization lifecycle, while others require controlled mutations or browser-visible confirmation. These properties lead to three practical challenges.

\begin{itemize}

\item \textit{Challenge 1: Layer ambiguity.}
A single session may contain local-client callbacks, remote MCP callbacks, and upstream authorization flows. Without distinguishing these layers, a detector may apply a rule to the wrong OAuth flow or miss flaws that only arise in delegated authorization.

\item \textit{Challenge 2: Lifecycle dependence.}
Several flaws cannot be determined from an isolated request. They only become visible after correlating authorization requests, callbacks, and token exchanges into a complete authorization lifecycle. Delegated flows further require this correlation across both layers.

\item \textit{Challenge 3: Confirmation boundary.}
Different flaws require different levels of evidence. Some can be checked passively from observed traffic, while others require carefully bounded request mutations. UI-dependent flaws, such as consent page bypass, require confirmation of what the user actually sees in the browser.

\end{itemize}

To address these challenges, we propose three design ideas.

\begin{itemize}

\item \textit{Solution idea 1: Layer-aware traffic identification.}
To address Challenge 1, the framework first extracts OAuth parameters and classifies callbacks into local-client and remote-server layers based on their \texttt{redirect\_uri} patterns. This step determines which OAuth flow a request belongs to before any flaw-specific rule is applied.

\item \textit{Solution idea 2: Lifecycle reconstruction.}
To address Challenge 2, the framework links authorization requests, callbacks, and token exchanges using \texttt{state} values and authorization codes, reconstructing complete OAuth lifecycles before applying flaw-specific checks. This makes it possible to reason about properties that span multiple messages, such as PKCE binding, state consistency, and code reuse.

\item \textit{Solution idea 3: Evidence-aware confirmation.}
To address Challenge 3, the framework separates passive checks, active probes, and UI-assisted validation. This lets us use low-impact passive rules where possible, while reserving controlled mutations and manual browser inspection for flaws that require stronger evidence.

\end{itemize}

Based on these ideas, we design a four-stage detection pipeline, shown in Figure~\ref{fig:auditing}. The first two stages build the context needed to interpret MCP OAuth traffic, and the last two stages apply flaw-specific checks with evidence levels matched to each flaw.

\subsection{Detection Pipeline}

The pipeline proceeds from context construction to vulnerability confirmation. It first identifies OAuth-related traffic and separates authorization layers, then reconstructs authorization lifecycles from the observed messages. Once this context is available, the framework applies passive checks that can be decided from traffic alone and active probes for flaws that require controlled interaction with the deployment.

\begin{figure*}[t]
    \centering
    \includegraphics[width=\linewidth]{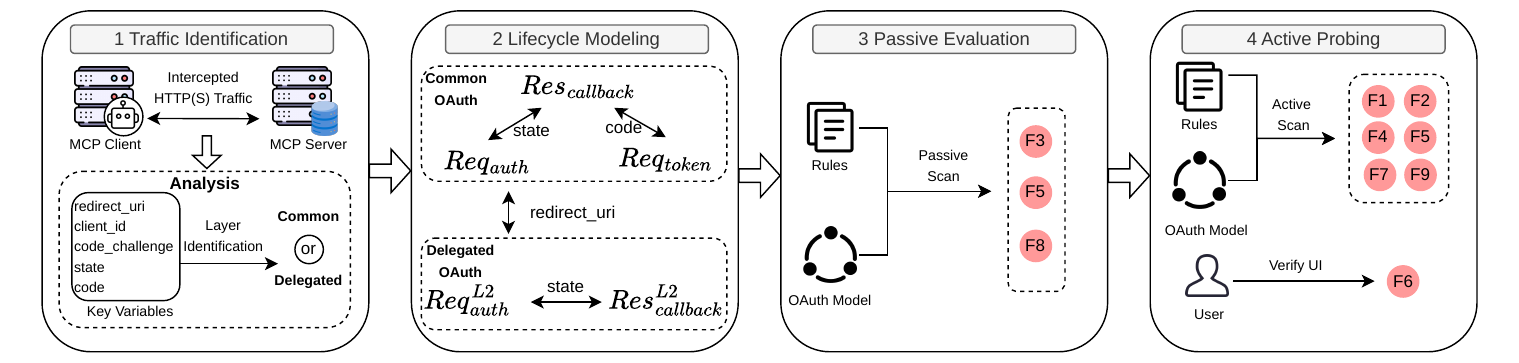}
    \caption{Detection pipeline of our framework for detecting flaws in remote MCP OAuth deployments.}
    \label{fig:auditing}
\end{figure*}

\textbf{(1) Traffic identification.}
The framework begins by filtering OAuth-related interactions from raw HTTP(S) traffic, isolating the authorization-relevant subset from background requests. This is done by extracting the key variables shown in Figure~\ref{fig:auditing}, including \texttt{redirect\_uri}, \texttt{client\_id}, \texttt{code\_challenge}, \texttt{state}, and authorization \texttt{code}. The framework then infers the authorization layer from the destination of the callback endpoint. Loopback addresses and custom URI schemes---such as \texttt{127.0.0.1} and \texttt{vscode://}---indicate that the callback targets the local MCP client layer (L1), whereas callbacks pointing to a remote MCP server URL indicate a delegated upstream authorization layer (L2). This layer inference separates ordinary single-layer OAuth flows from multi-layer delegated MCP flows, ensuring that layer-specific detection rules are applied only to the appropriate authorization context.

\textbf{(2) Lifecycle modeling.}
The framework abstracts each standard OAuth lifecycle into three core components: the authorization request sent from the client to the authorization server, the redirect callback carrying the authorization code returned by the server, and the token exchange request that redeems the code for an access token. Correlating these three messages across a single flow is essential for detecting flaws that span multiple protocol steps. The shared \texttt{state} parameter serves as the binding key that links the authorization request to its corresponding callback, while the authorization \texttt{code} value links the callback to the subsequent token exchange. For multi-layer delegated authorization, the framework additionally reconstructs the second-layer (upstream service layer) authorization request and callback pair, and records the \texttt{redirect\_uri} or routing context that bridges the two authorization layers. Together, these linked structures provide the complete per-flow context against which passive and active detection steps operate, corresponding to the standard and delegated branches shown in Figure~\ref{fig:auditing}.

\textbf{(3) Passive evaluation.}
Against the reconstructed lifecycle, the framework applies passive rules that do not send additional traffic. These rules include:
\begin{itemize}
    \item \textit{F3 (Layer Inconsistency):} Check whether PKCE is consistently enforced across L1 and L2.
    \item \textit{F5 (PKCE Downgrade):} Check whether authorization requests include valid PKCE parameters and avoid the \texttt{plain} method.
    \item \textit{F8 (Weak state):} Check whether \texttt{state} is present and sufficiently unpredictable for CSRF protection.
\end{itemize}
Passive evaluation gives a low-impact baseline before any active mutation is attempted.

\textbf{(4) Active probing.}
For flaws that cannot be confirmed through passive analysis alone, the framework executes targeted active probes:
\begin{itemize}
    \item \textit{F1 (Malicious DCR):} Submit a DCR request containing a malicious \texttt{redirect\_uri} to test whether the server enforces proper boundary controls.
    \item \textit{F2 (Blind Client Trust):} Replaces the \texttt{client\_id} with a spoofed identifier (e.g., \texttt{evil\_client\_id}) to test whether the system accepts unregistered identities.
    \item \textit{F4 (Nested Context Pollution):} Decodes and tampers with the nested \texttt{redirect\_uri} within the \texttt{state} parameter.
    \item \textit{F5 (PKCE Downgrade):} Performs PKCE downgrade testing: changing \texttt{code\_challenge\_method} from \texttt{S256} to \texttt{plain}, or stripping both \texttt{code\_challenge} and \texttt{method} entirely.
    \item \textit{F7 (Open Redirect):} Mutates the \texttt{redirect\_uri} to an attacker-controlled malicious address.
    \item \textit{F9 (Code Replay):} Intercepts and replays a consumed authorization code (\texttt{code}) to test single-use enforcement.
\end{itemize}

F6 requires a different treatment because it depends on what the browser presents to the user. For this flaw, testers manually trigger generated test links, including truncated mid-flow URLs, and inspect whether the consent page, final \texttt{redirect\_uri}, and risk warnings are displayed in the expected order.

The framework uses three levels of evidence. F3 and F8 are detected passively from reconstructed lifecycles. F5 combines passive checks of observed PKCE parameters with active downgrade probes. F1, F2, F4, F7, and F9 require active probes followed by manual confirmation. F6 is UI-assisted: the framework generates test links, while researchers confirm the browser-visible behavior. All vulnerable cases reported below were manually verified.

\textbf{Implementation Details}. 
We use VSCode Copilot as the MCP client and simultaneously leverage Burp Suite~\cite{burp} to monitor its communication traffic. Our detection framework runs as a Burp plugin, performing vulnerability checks on the captured requests and responses.
Our automated detection framework is built upon and extends OAuthScan, an existing Burp Suite extension~\cite{oauthscan}. The OAuthScan is capable of identifying basic OAuth flaws (e.g., open redirects and weak state parameters). However, it lacks the architectural context required in MCP environments. Consequently, we restructured its core logic and extended its capabilities to address the three key characteristics of MCP identified earlier. Specifically, we introduced customized lifecycle modeling. We overhauled its traffic classification rules to support context correlation across delegated authorization layers (L1/L2), and integrated new detection logic that also applies to common OAuth flaws.

\subsection{Detection Results}

\textbf{Dataset and Performance.}
We use the 119 testable DCR-enabled OAuth servers from Table~\ref{tab:oauth-subset} as the evaluation dataset and apply the detection pipeline to identify candidate flaw instances. Each candidate alert is then manually verified before being counted in the results below. For passive detections, researchers re-inspected the relevant traffic to confirm missing, inconsistent, or downgraded parameters. For active probes, researchers checked whether the server accepted the mutated request and whether the observed behavior matched the flaw definition. Consistent with our ethical constraints, we did not complete exploit chains that would access, modify, or exfiltrate real user data.

\begin{table}[!h]
\centering
\caption{Confusion matrix of the detection framework after manual verification.}
\label{tab:confusion-matrix}
\small
\renewcommand{\arraystretch}{1.2}
\begin{tabular}{l c c}
\hline
\textbf{Detection outcome} & \textbf{Vulnerable} & \textbf{Not vulnerable} \\
\hline
Flagged by tool & 325 (TP) & 54 (FP) \\
Not flagged by tool & 1 (FN) & N/A \\
\hline
\end{tabular}
\end{table}

Table~\ref{tab:confusion-matrix} summarizes the tool-level detection outcome after manual verification. The framework produced 379 candidate alerts, of which 325 were confirmed as true positives, and 54 were false positives. We also identified 1 false negative during manual review. This corresponds to 85.75\% precision and 99.69\% recall over the manually verified flaw instances. We do not report true negatives because the detector operates over flaw-specific candidate opportunities rather than an exhaustively enumerated set of all non-vulnerable request variants.

The false positives mainly stem from two engineering constraints. The first is deep redirect truncation, accounting for 18 cases in F2 and 25 cases in F7. To preserve scanning efficiency, the tool restricts HTTP redirect tracking to a default maximum of five hops, which can miss validations performed deeper in the redirect chain. The second is background traffic noise, involving 7 cases in F3 and 4 cases in F8, where unrelated external OAuth flows were conflated with the target authorization flow. The single false negative appeared in F5 and resulted from non-standard parameter naming that evaded our baseline signature matching.

\begin{figure}[!h]
\centering
\includegraphics[width=0.5\textwidth]{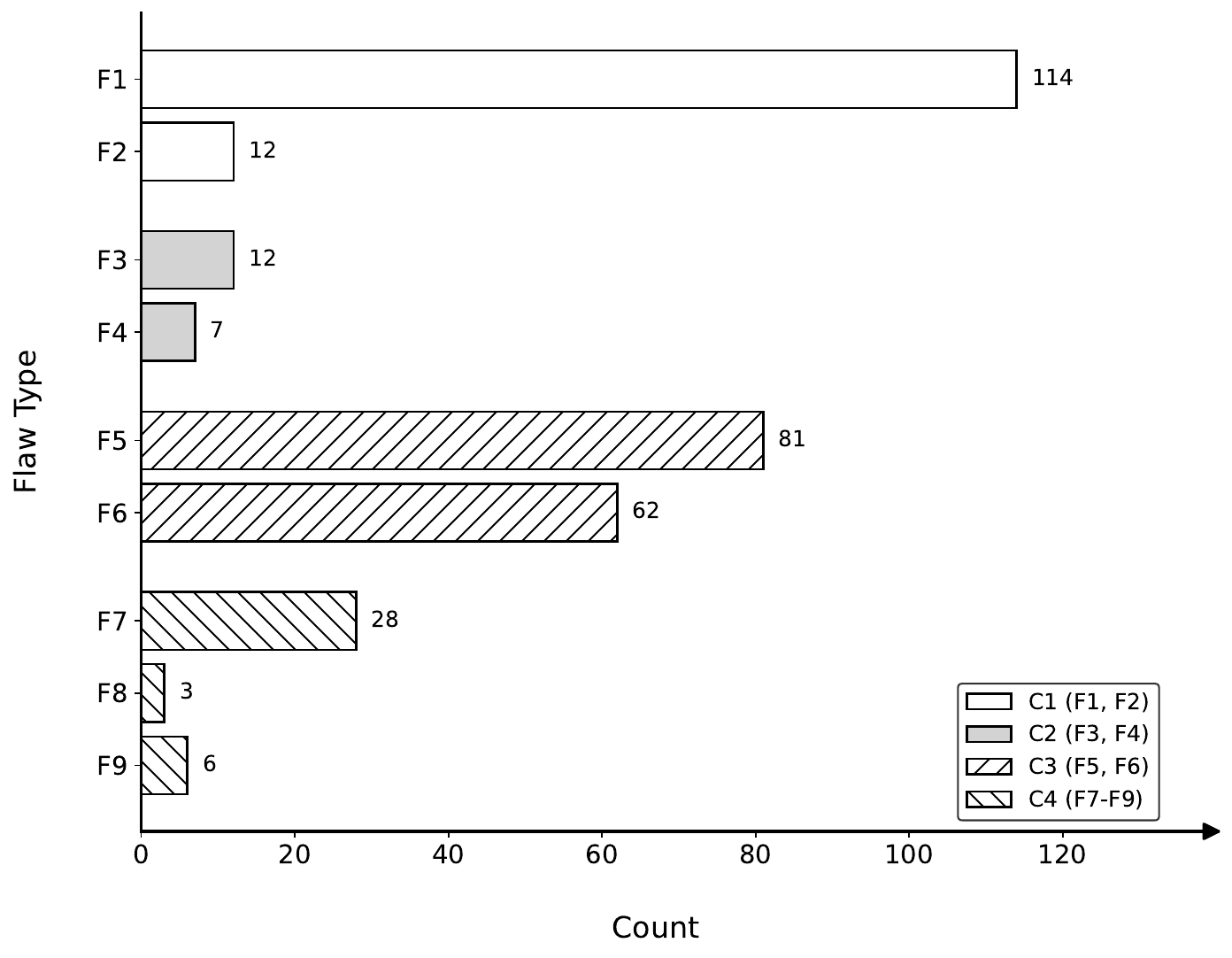}

\caption{Overall flaw detection results on 119 MCP servers.}
\label{fig:flaw-dist}
\end{figure}

\textbf{Finding 3.1: Every server in our evaluation dataset exhibits at least one confirmed authentication flaw.}
All 119 servers exhibited at least one confirmed flaw across the four categories, yielding 325 confirmed flaw instances in total, reflecting the pervasive presence of authentication weaknesses in real-world MCP deployments.

More strikingly, 39 servers (32.8\%) were flagged by three or more categories, indicating that authentication weaknesses often appear in combination rather than as isolated mistakes.
Figure~\ref{fig:flaw-dist} shows the per-flaw detection results.

\textbf{Finding 3.2: MCP-specific categories dominate the evaluation dataset, led by Dynamic Client Registration Flaws (C1) and Open Client Environment Flaws (C3).}
115 of 119 servers (96.6\%) exhibited at least one C1 flaw.
Within this category, F1 (Malicious DCR Binding) is the dominant flaw, confirmed in 114 servers (95.8\%): their DCR endpoints accept any \texttt{redirect\_uri} submitted by an anonymous registrant, allowing an attacker to register a malicious callback and intercept authorization codes. F2 (Blind Client Trust) was confirmed in 12 servers (10.1\%), where the authorization server accepted spoofed \texttt{client\_id} values without verifying registration status.

102 of 119 servers (85.7\%) exhibited at least one C3 flaw.
F5 (PKCE Downgrade) was confirmed in 81 servers (68.1\%): their authorization servers either allow the omission of \texttt{code\_challenge} or accept requests with the \texttt{code\_challenge\_method} downgraded to \texttt{plain}, nullifying PKCE protection even when nominally supported.

F6 (Consent Page Bypass) was confirmed in 72 servers (60.5\%): failing to display the \texttt{redirect\_uri}, allowing attackers to deceive users into approving requests that leak authorization codes to rogue localhost ports.

\textbf{Finding 3.3: Delegated authorization is common in the evaluation dataset, and its multi-hop structure introduces cross-layer inconsistencies.}
81 of 119 servers (68.1\%) implement delegated authorization (PA), and among these, 40 (49.4\%) use nested \texttt{state} parameters to pass routing context across authorization layers, which is the very mechanism exploited by F4. 

Furthermore, 18 servers (15.1\%) exhibited at least one C2 flaw: 12 confirmed instances of F3 (Layer Inconsistency), where PKCE is enforced at layer 1 but omitted at the upstream layer 2 request, and 7 instances of F4 (Nested Context Pollution), where the embedded \texttt{redirect\_uri} inside the \texttt{state} parameter is accepted without integrity verification or whitelist validation.

\textbf{Finding 3.4: Common OAuth Misconfigurations (C4) persist at lower but non-negligible rates.}
34 of 119 servers (28.6\%) exhibited at least one C4 flaw. 
F7 (Open Redirect) was confirmed in 28 servers (23.5\%).
Among them, 15 accept fully substituted attacker-controlled domains, while 13 exhibit weaker forms (e.g., accepting decimal IP representations or non-existent subpaths). 
F9 (Code Replay) was confirmed in 6 servers (5.0\%), and F8 (Weak State) in 3 servers (2.5\%).

For confirmed flaws with practical security impact, we initiated responsible disclosure to affected vendors. Each report included the vulnerability principle, reproduction steps, and mitigation suggestions tailored to the corresponding flaw type. Several vendors have acknowledged or confirmed our reports, and 9 confirmed vulnerabilities have been assigned CVE IDs.

\begin{findingbox}
\textbf{Finding 3.} Authentication flaws are prevalent in the testable OAuth-enabled MCP servers, and flaw categories tied to MCP-specific patterns, especially dynamic client registration and open client environments, appear most frequently.
\end{findingbox}

\subsection{Case Studies}

We select three representative cases to cover the three MCP-specific deployment characteristics and to show how individual flaws compose into end-to-end account takeover chains. 
Note that all cases discussed in this section have been responsibly disclosed to the respective vendors.

\begin{figure}[h]
\centering
\includegraphics[width=\linewidth]{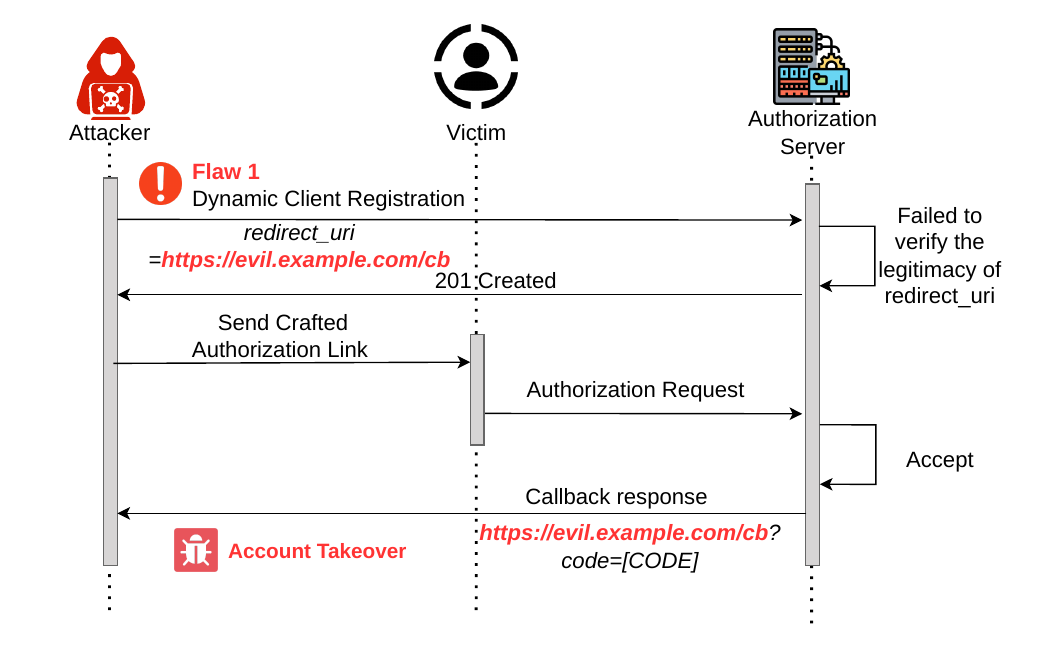}

\caption{Case Study 1: Malicious client registration via open DCR.}
\label{fig:casestudy1}
\end{figure}

\textbf{Case Study 1: Malicious client registration via open DCR.} 
When a server permits open DCR with arbitrary callback URLs, attackers can exploit this registration phase as a malicious routing mechanism to hijack authorization codes.
Figure~\ref{fig:casestudy1} shows the attack flow: the attacker first sends a registration request with an attacker-controlled \texttt{redirect\_uri}, crafts a malicious authorization URL containing this URI, and induces the victim to interact with it, causing the authorization code to be delivered to the attacker's endpoint.

\textit{Case-1: **\footnotemark
\footnotetext{Anonymized for ethical considerations.}(https://mcp.**.dev/mcp)}. This MCP server enables AI coding tools to directly access and query application error monitoring data, thereby automatically analyzing root causes and assisting with fixes. This server is representative of the CVE-2026-26384 to CVE-2026-26390 series and exposed a public \texttt{registration\_endpoint} in its \texttt{.well-known/\allowbreak oauth-authorization-server} metadata. The attacker can issue a DCR request with an attacker-controlled \texttt{redirect\_uri} (e.g., \texttt{https://evil.example.com/cb}) and receive a legitimately issued \texttt{client\_id}. Using this ID, the attacker constructs an authorization URL that contains the attacker-controlled \texttt{redirect\_uri} and social-engineers the victim into clicking it. After the victim completes the consent flow, the authorization server delivers the authorization code to the attacker's server, allowing the attacker to exchange it for an access token and take over the victim's MCP session. We reported this vulnerability and obtained CVE-2026-26390.

\begin{figure}[h]
\centering
\includegraphics[width=\linewidth]{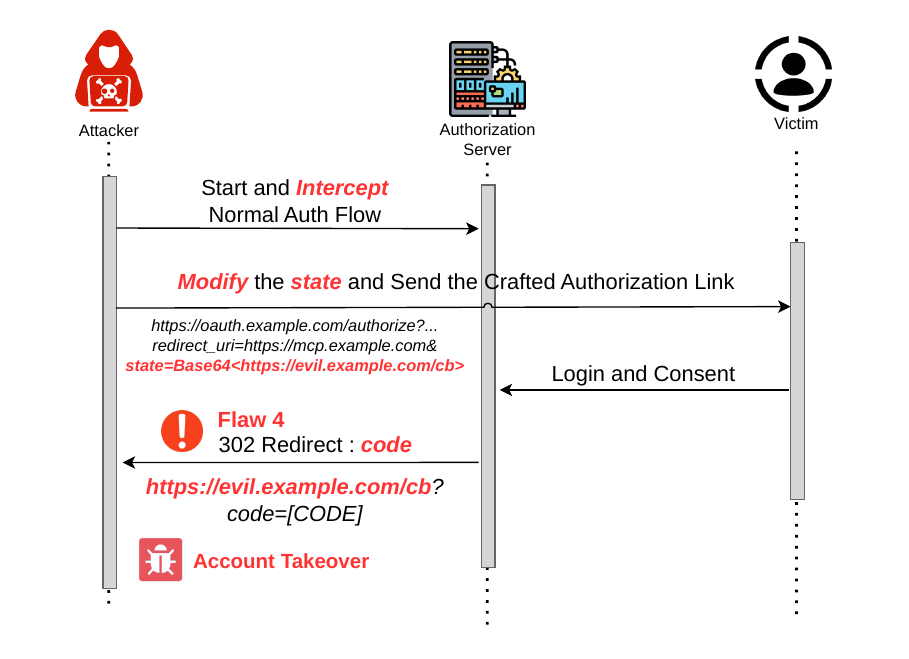}

\caption{Case Study 2: Nested context pollution leading to account takeover.}
\label{fig:casestudy2}
\end{figure}

\textbf{Case Study 2: Nested context pollution leading to account takeover.} Delegated authorization introduces a second authorization context between the MCP server and an upstream service. If the MCP server serializes downstream routing context into client-visible OAuth state without integrity protection, an attacker can tamper with that context and redirect a valid upstream authorization result to an attacker-controlled endpoint. Figure~\ref{fig:casestudy2} illustrates this delegated attack pattern.

\textit{Case-2: **\footnotemark[\value{footnote}] (https://mcp.**.tech/mcp).} 
This MCP server provides tools for managing projects, branches, queries, and database migrations. In this delegated authorization architecture, the MCP server exposed a critical nested context pollution vulnerability (F4). The attacker first obtains a legitimate L2 authorization request generated by the MCP server, decodes the \texttt{state} parameter, and mutates its nested \texttt{redirect\_uri} field to an attacker-controlled domain (e.g., \texttt{https://evil.example.com/cb}) before re-encoding it. When the victim clicks the forged link and completes authorization at the upstream identity provider, the callback carrying the valid authorization code is sent back to the MCP server. Because the MCP server fails to enforce integrity validation on the nested \texttt{state} payload, it blindly parses the tampered \texttt{redirect\_uri} and executes a secondary redirect. This forwards the victim's authorization credentials directly to the attacker's server, resulting in account takeover. 

\begin{figure}[h]
\centering
\includegraphics[width=\linewidth]{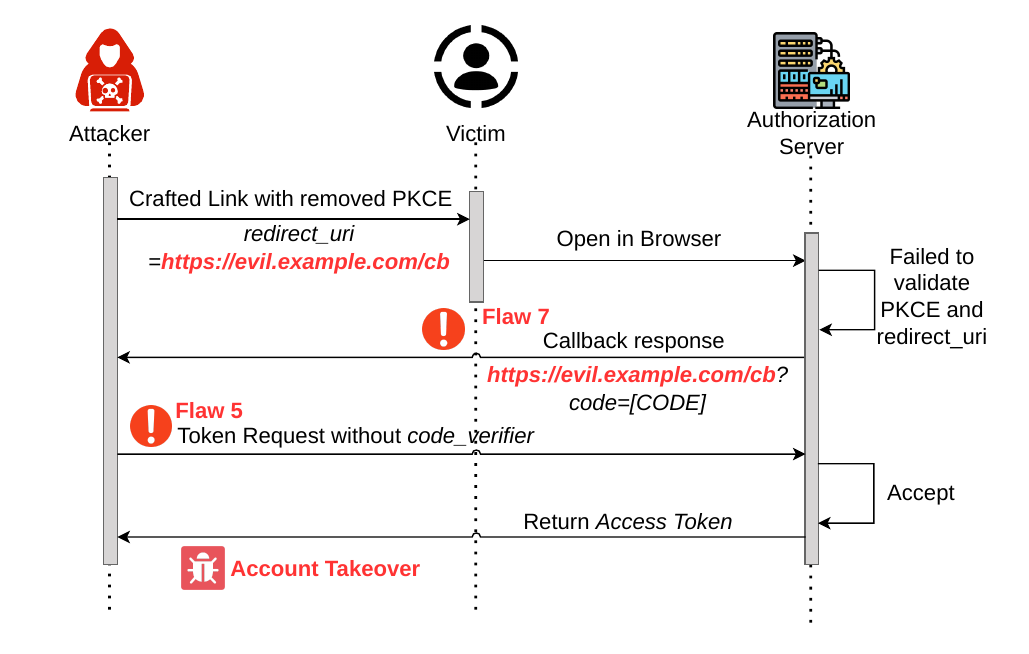}

\caption{Case Study 3: Open redirect amplified by PKCE downgrade.}
\label{fig:casestudy3}
\end{figure}

\textbf{Case Study 3: Open redirect amplified by PKCE downgrade.} Open-client MCP deployments rely heavily on PKCE because local clients cannot safely keep long-term secrets. When an authorization server both accepts an attacker-controlled \texttt{redirect\_uri} and allows PKCE to be omitted or downgraded, stealing an authorization code becomes sufficient for token theft. Figure~\ref{fig:casestudy3} shows this composition.

\textit{Case-3: **\footnotemark[\value{footnote}] (https://mcp.**.com/mcp).}
This MCP server enables AI tools to perform create, read, update, and delete operations on documents and projects within the ** workspace. This server concurrently exposed an open redirect (F7) and a lack of mandatory PKCE enforcement (F5). The attacker first crafts an authorization URL that mutates the \texttt{redirect\_uri} to an attacker-controlled domain (e.g., \texttt{https://evil.example.com/cb}) and removes the \texttt{code\_challenge} and  \texttt{code\_challenge\_method} parameters. When the victim authorizes the request, the server issues a valid authorization code and redirects it to the attacker-controlled endpoint. Because no PKCE challenge was recorded during code issuance, the attacker can exchange the code without a \texttt{code\_verifier}, obtaining the victim's access token and achieving account takeover. We reported this vulnerability and obtained CVE-2025-69898.

\begin{findingbox}
\textbf{Finding 4.} Confirmed flaws can lead to high-impact attacks: weak client registration, unsafe delegated context handling, and redirect or PKCE weaknesses can cause authorization-code or token leakage, resulting in sensitive information exposure and account takeover.
\end{findingbox}

\section{Discussion}

\subsection{Root Cause Analysis}

The flaws we observe are not merely isolated implementation mistakes; they are amplified by the mismatch between a rapidly evolving specification and fast-moving real-world deployments. Within roughly one year, MCP moved from having no mandatory authentication requirements to adopting a more complete OAuth-based authorization framework. Many deployments appear to have implemented only the minimum flow needed for interoperability, while leaving security-critical checks such as PKCE enforcement, redirect URI validation, and client registration controls incomplete. The fact that 40.55\% of discovered servers still expose tools without authentication further suggests that authentication is often treated as a deployment add-on rather than a first-class security boundary.

A second root cause lies in the intermediary role of MCP servers. In delegated deployments, the server is not simply a protected resource; it also becomes an OAuth client to upstream services. This creates a multi-hop authorization chain in which security properties must be preserved across independently operated systems. 
Nested routing state and inconsistent protection between local and upstream hops are all symptoms of this design pressure. These failures are therefore better understood as coordination failures across layers than as ordinary single-endpoint misconfigurations.

Dynamic client registration introduces a third and particularly important source of risk. Although the specification treats DCR as an optional fallback, our measurement shows that all 1,118 OAuth-enabled servers advertise a \texttt{registration\_endpoint}, and seven of the nine CVEs we obtained correspond to F1. This suggests that developers frequently default to DCR because it is convenient and readily available in off-the-shelf OAuth implementations, but deploy it without strong redirect URI restrictions or client identity checks. In practice, DCR becomes the easiest path to interoperability and, at the same time, the broadest attack surface.

\subsection{Mitigation Suggestions}

Based on our findings, we distill concrete mitigations at three levels: authorization server implementation, MCP server deployment, and specification hardening. The suggestions below are organized by the flaw categories they most directly address.

\textit{Restrict client registration.}
Authorization servers should treat DCR endpoints as semi-trusted interfaces rather than open APIs. Concretely: (1)~enforce an allowlist of permitted \texttt{redirect\_uri} patterns (e.g., rejecting arbitrary internet-routable domains); (2)~require client attestation or rate-limit registrations per IP; and (3)~migrate from DCR to the Client ID Metadata Document (CIMD) mechanism introduced in the 2025-11-25 MCP specification, which pins client identity to a cryptographically verifiable HTTPS-hosted JSON document rather than an open registration call.

\textit{Enforce PKCE server-side.}
PKCE enforcement is a server-side responsibility. Authorization servers must reject authorization requests that omit \texttt{code\_challenge}, and must not accept \texttt{plain} as a valid \texttt{code\_challenge\_method}. The server's metadata document should advertise only \texttt{S256} in \texttt{code\_challenge\_methods\_supported}. Given that 67.5\% of tested servers silently accept PKCE-free requests, this single change would neutralize the most prevalent single flaw in our dataset.

\textit{Preserve user-visible consent.} To mitigate localhost impersonation, authorization servers must enforce strict UI transparency by unconditionally displaying the exact \texttt{redirect\_uri} on the consent page. 
Furthermore, servers should treat \texttt{localhost} callbacks with elevated scrutiny, ideally presenting visual warnings to ensure users explicitly acknowledge the exact callback destination before granting authorization.

\textit{Isolate delegated contexts.}
MCP servers implementing delegated authorization must not embed routing parameters (e.g., downstream \texttt{redirect\_uri}) inside the \texttt{state} parameter without integrity protection. Instead, the MCP server should maintain a server-side mapping from an opaque \texttt{state} parameter to the routing context, preventing client-side tampering. 

\textit{Harden the specification defaults.}
The MCP specification should elevate PKCE enforcement and DCR redirect-URI restrictions from SHOULD to MUST-level requirements. The current MAY-level status of CIMD as the preferred registration mechanism should be strengthened to RECOMMENDED over DCR, with DCR explicitly classified as a high-risk option requiring additional safeguards.


\subsection{Limitations and Future Work}

Our study has several limitations. First, our measurement relies on FOFA and Shodan for initial discovery, which may not capture MCP servers deployed behind CDNs, firewalls, or private networks, introducing coverage bias toward publicly indexed infrastructure; this scope was also guided by ethical considerations, as we restricted our study to publicly reachable assets to minimize risk to deployed systems and operators. Second, our flaw detection framework relies on rule-based matching, which requires manual specification of each flaw type and may miss subtle or novel vulnerability patterns.
Future work could replace this with LLM or agent-driven analysis for more adaptive and comprehensive detection. 

As AI agents increasingly interact with external services on behalf of users, the authentication patterns established by MCP will likely influence other agent protocols such as A2A~\cite{a2a} and ANP~\cite{anbiaee2026security}. Studying whether the same flaw classes recur in those ecosystems and developing standardized security testing methodologies for agent authentication layers more broadly remain important open problems.

\section{Related Work}
\label{sec:relatedwork}

\textbf{MCP security.}
Existing MCP security research has mainly studied model-layer and tool-interface threats. Hou et al.~\cite{hou2025mcp} defined a 16-threat taxonomy across four attacker types, Gaire et al.~\cite{gaire2025sok} systematized MCP security and safety knowledge, and Anbiaee et al.~\cite{anbiaee2026security} compared MCP with A2A, Agora, and ANP. Empirical studies have reported tool-layer flaws, including tool poisoning in 5.5\% of 1,899 servers~\cite{hasan2025mcp}, server hijacking risks in a substantial portion of 67,057 analyzed registry entries~\cite{ligao2025mcp}, and attacks that exploit agents' reliance on tool descriptions~\cite{guo2025mcp}. Other work studies MCP toolchain attacks~\cite{zhao2025parasitic}, malicious server taxonomies~\cite{zhao2025taxonomy}, clause-compliance vulnerabilities~\cite{yang2026compatibility}, defensive frameworks~\cite{narajala2025enterprise, errico2025securing}, and OAuth-enhanced tool definitions~\cite{bhatt2025etdi}. Closest to our work, Huang et al.~\cite{huang2026caller} identified caller identity confusion in MCP deployments, showing that servers commonly reuse cached authorization state across tool invocations regardless of caller identity. Prakash~\cite{prakash2026aip} proposed the Agent Identity Protocol after observing widespread unauthenticated servers. These studies motivate stronger authentication, while our work analyzes OAuth implementation-layer flaws in remote MCP deployments.

\textbf{Agent authorization security.}
Recent work on agentic authorization highlights open problems in delegation and scope control. South et al.~\cite{south2025delegation} proposed authenticated delegation frameworks for AI agents, extending OAuth 2.0 and OpenID Connect with agent-specific credentials and metadata to enforce scoped, auditable delegation chains. 
The OpenID Foundation~\cite{openid2025agentic} identified recursive delegation and scope attenuation as unresolved challenges, and noted the absence of standard mechanisms for agents acting across service boundaries, including DCR and asynchronous authorization. These works address delegation at the protocol design level.
Our work instead measures how OAuth is concretely implemented in deployed MCP servers and characterizes the implementation-layer flaws that arise from MCP's specific architectural characteristics.

\textbf{OAuth security.}
OAuth security has been studied from formal, empirical, and attack perspectives. Formal analyses by Fett et al.~\cite{fett2016oauth, fett2017oidc, fett2019fapi} established security guarantees for OAuth 2.0 and OpenID Connect, while uncovering previously unknown attacks. 
Hosseyni et al.~\cite{hosseyni2025audience} further discovered audience injection attacks across a broad family of OAuth-derived protocols, directly relevant to MCP's delegated setting. On the empirical side, OAuch~\cite{philippaerts2022oauch} found that 97 of 100 OAuth IdPs leave at least one threat unmitigated, with PKCE downgrade succeeding against 43\% of PKCE-supporting providers; follow-up work confirmed persistent non-compliance~\cite{philippaerts2023revisiting}, and further studies documented OAuth failures on web~\cite{yang2016model, zhou2014ssoscan} and mobile platforms~\cite{wang2015vulnerability, wang2016achilles}. Known attack techniques include redirect URI manipulation~\cite{wang2019redirect, innocenti2023redirect, khodayari2025redirect}, cross-app OAuth attacks in integration platforms~\cite{luo2025crossapp}, and redirect chain injection in SSO brokers~\cite{innocenti2025broker}.

Our work builds on these foundations but focuses on how these flaws interact with MCP-specific deployment characteristics, including open client environments, dynamic client registration, and delegated authorization. 
These characteristics are structurally distinct from the prior OAuth research, which amplify known OAuth weaknesses and introduce new attack surfaces absent in the general Web and mobile OAuth settings studied by prior work.

\section{Conclusion}

This paper presented the first measurement study of authentication security in remote MCP servers. We identified approximately 7,973 live deployments, found that 40.55\% expose tool interfaces without authentication, and showed that OAuth-enabled deployments commonly combine open client environments, dynamic client registration, and delegated authorization. 
From these characteristics, we derived a taxonomy of 9 flaws across 4 categories and built a semi-automated detection framework to evaluate real deployments. 
Applying it to 119 testable OAuth-enabled servers, we found that every server exhibited at least one flaw, with dominant categories rooted in MCP-specific deployment patterns, and obtained 9 CVE IDs through responsible disclosure. These results show that authentication weaknesses in remote MCP are an emerging protocol-infrastructure risk for agentic ecosystems.

\section*{Ethics Considerations}

Our study strictly followed academic research norms and ethical guidelines. Because this work involved discovering and validating authentication vulnerabilities in real-world remote MCP deployments, we took the following measures to minimize risk to deployed systems, service operators, and users.

\textit{Publicly reachable assets only.}
We restricted discovery to assets indexed by mainstream cybersecurity search engines and interacted only with endpoints that were already exposed on the public Internet. We did not attempt to bypass access controls, scan private address spaces, or test enterprise-internal assets that were not publicly reachable.

\textit{Low-impact validation with our own accounts.}
Our active verification was limited to the minimum protocol actions needed to confirm authentication flaws, such as fetching metadata documents, issuing MCP initialization requests, testing whether dynamic client registration accepted attacker-controlled \texttt{redirect\_uri} values, checking PKCE enforcement, mutating \texttt{state} or \texttt{redirect\_uri} parameters in authorization requests, and replaying authorization codes only in tester-controlled sessions to check single-use enforcement. 
For flaws requiring user-interface validation, such as consent-page bypass, testers manually inspected the authorization flow using their own accounts and browser sessions. 
We did not stress servers, persist access, perform destructive operations, invoke sensitive MCP tools at scale, or use issued tokens to access, modify, or exfiltrate real user data.
When a full attack chain would have required compromising an actual third-party account or completing a real account takeover, we stopped after confirming the vulnerable condition and validated the end-to-end impact only in controlled environments.

\textit{Responsible disclosure.}
All confirmed vulnerabilities were reported to affected vendors through responsible disclosure channels, including direct email and vulnerability platforms where available. This process resulted in 9 assigned CVE IDs at the time of submission.

\bibliographystyle{IEEEtran}
\bibliography{ref}

\end{document}